\documentclass[reprint,amsmath,amssymb,aip,prl]{revtex4-2}
\usepackage{graphicx}
\usepackage{dcolumn}
\usepackage{bm}
\usepackage[mathlines]{lineno}

\DeclareMathAlphabet{\mathcal}{OMS}{cmsy}{m}{n}
\DeclareSymbolFont{largesymbols}{OMX}{cmex}{m}{n}

\usepackage[bookmarks=true,colorlinks,linkcolor=blue,urlcolor=blue,citecolor=blue,plainpages=false,pdfpagelabels,final,breaklinks=true]{hyperref}

\usepackage{marvosym}
\usepackage{hyperref}

\usepackage{fancyhdr}
\usepackage{amsmath}
\usepackage{autobreak}
\allowdisplaybreaks

\usepackage{ragged2e}

\usepackage{geometry}
\begin{document}

\preprint{APS/123-QED}

\title{Dynamics of Magnetic Skyrmions under Temperature Gradients}

\author{Chaofan Gong}
\affiliation{Department of Physics,The Chinese University of HongKong, Hong Kong, China}

\author{Yan Zhou}
\affiliation{School of Science and Engineering, The Chinese University of Hong Kong, Shenzhen, Guangdong, 518172, China}

\author{Guoping Zhao}
\affiliation{College of Physics and Electronic Engineering, and Center for Computational Sciences, Sichuan Normal University, Chengdu, 
610068, China}

\begin{abstract}
Authors to whom correspondence should be addressed: [Chaofan Gong, chaofangong.physics@gmail.com; Yan Zhou, zhouyan@cuhk.edu.cn]

We expand the Zhang-Li spin-transfer torque [Phys. Rev. Lett. 93, 127204 (2004)] to finite temperatures by the scattering amplitude.
Considering various factors including the adiabatic and diabatic effects of electrons and magnons, entropy equivalent field, 
thermal dipole field, thermal activation, magnetic-anisotropy gradient, and diffusion, we answer a recent question both experimentally 
and theoretically controversial: Will skyrmions move to the hot or cold region under thermal gradients?
\end{abstract}

\maketitle
\thispagestyle{fancy}

Magnetic skyrmions in the system with the Dzyaloshinsky-Moriya interaction (DMI) are nontrivial spin textures featured by the 
topological charge $Q=(1/4\pi)\int{(\boldsymbol{m}\cdot \partial_x\boldsymbol{m}\times \partial _y\boldsymbol{m}) dxdy}$ 
($\boldsymbol{m}$ is the unit magnetization vector in the $x$-$y$ plane) \cite{muhlbauer2009Skyrmion,nagaosa2013Topological}.
They have the emergent electromagnetic field \cite{schulz2012Emergent} deflecting particle flows, resulting in the skyrmion Hall effect
described by the Hall angle $\theta =\arctan (v_{\parallel}/v_{\bot})$,
where $v_{\parallel}$ and $v_{\bot}$ represent the skyrmion velocity parallel and perpendicular to the driving force, respectively.
So far, skyrmions are conveniently manipulated by various means, such as optics \cite{wang2020Optical}, 
electricity \cite{litzius2020Role}, magnetism \cite{zhang2018Manipulation}, heat \cite{lin2014Ac}, and stress 
\cite{gusev2020Manipulation}.
Among them, the temperature gradient demonstrates a promising and efficient means of collecting waste heat for energy-efficient green 
spintronics.

However, under temperature gradients, experiments show a controversial result that skyrmions can move toward the 
cold \cite{wang2020Thermala} or hot \cite{yu2021Realspace} region.
This is contrary to most theories \cite{kong2013Dynamics,kovalev2014Skyrmionic,lin2014Ac,liu2016Controla,weissenhofer2021Skyrmion}
that skyrmions should move from the cold to the hot region.
Only one theory \cite{wang2021Rectilinear} based on the Ginzburg Landau equation shows that skyrmions can move towards the cold region, 
but it can not quantitatively explain all the experimental phenomena.

To address this conundrum, we develop the Zhang-Li spin-transfer moment \cite{zhang2004Roles} to finite 
temperatures by the scattering amplitude \cite{iwasaki2014Theory,schutte2014Magnonskyrmionb}.
Poring the entropy equivalent field \cite{kim2015LandauLifshitz,schlickeiser2014Role,wang2014Thermodynamica,
yan2015Thermodynamica}, heating induced dipole field \cite{moretti2017Domainc,shokr2019Steering}, thermal activation, and diffusion,
we present a possible answer to the lingering issue by showing that the skyrmion velocity can be controlled and even reversed by 
thermal gradients. 
Namely, the skyrmion velocity and trajectory can be manipulated by various experimental parameters such as the temperature, heating 
source size, film thickness, and Gilbert damping. 
Our theory can explain other key experimental findings such as the skyrmion Hall angle is dependent \cite{yu2021Realspace} 
or independent \cite{litzius2020Role} on the temperature, and the tangential velocity (i.e. $v_\parallel$) can be greater than the 
particle-flow direction (i.e. $v_{\bot}$) \cite{mochizuki2014Thermally}.

\begin{figure}[t]
\includegraphics[clip=true,width=1\columnwidth]{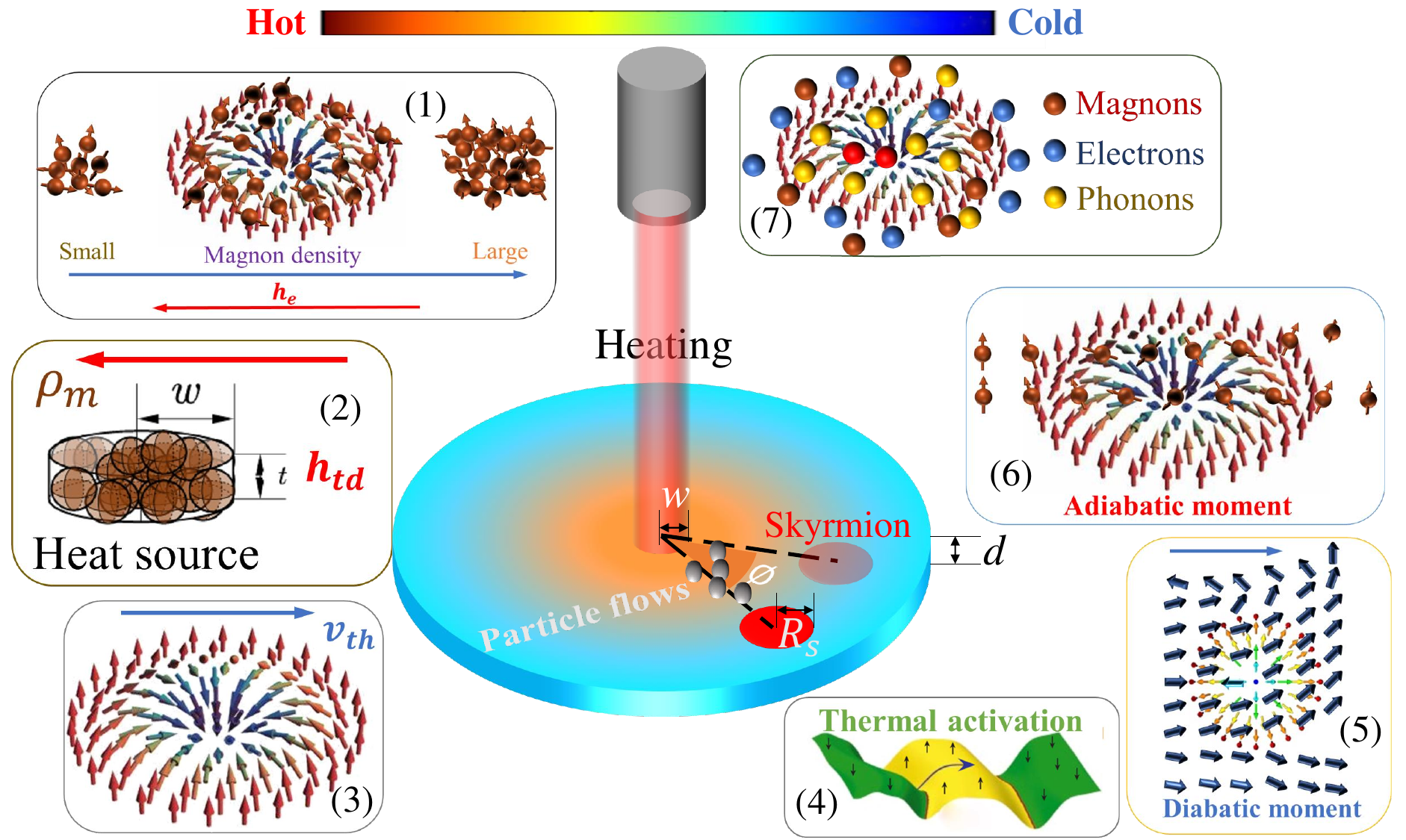}
\caption{The schematic diagram of skyrmion dynamics under thermal gradients (the laser heating here is for illustration only).
(1) The equivalent field produced by the magnon entropy.
(2) The dipole field produced by the magnetic charge.  
(3) The thermal diffusion.  
(4) The thermal activation. 
(5) The diabatic scattering. 
(6) The adiabatic spin-transfer torque. 
(7) Skyrmions in the heat bath lead to the equilibrium-magnetization reduction. 
}\label{Fig1}
\end{figure}

Under thermal gradients, skyrmions are in the heat bath of thermally excited particles (magnons, electrons, and phonons) 
\cite{weissenhofer2021Skyrmion}, as shown in Fig.\ref{Fig1}. 
The interaction between these particles and skyrmions makes the system disordered and effectively reduces the equilibrium magnetization 
$m_e(T)$ \cite{moreno2016Temperaturedependenta}. 
In fact, various interaction parameters including DMI, magnetocrystalline anisotropy, and Heisenberg exchange all depend on temperatures 
\cite{schlotter2018Temperaturea}, and thus the effective field components in the Landau-Lifshitz-Gilbert (LLG) equation at finite 
temperatures should be modified as the scaled equilibrium magnetization with a power law of $m_e\left(T\right)=(1-T/T_c)^x$ 
\cite{schlotter2018Temperaturea} ($x$ is the Bloch constant, and $T_c$ is the Curie temperature). 

To begin with, we realize that the electron transport mechanism can be extended to magnons \cite{mook2018Taking,lin2014Ac},
and thus the LLG equation can be popularized in the following generic form:
\begin{align}
\partial _t\boldsymbol{m}=&-\gamma \boldsymbol{m}\times \boldsymbol{h}_{eff}+\alpha \boldsymbol{m}\times \partial _t\boldsymbol{m}+
\sum_i{a_i\left( \boldsymbol{\mathcal{V}}_i \cdot \nabla \right) \boldsymbol{m}}   \nonumber \\
&-\sum_i{b_i\boldsymbol{m}\times \left( \boldsymbol{\mathcal{V}}_i \cdot \nabla \right) \boldsymbol{m}},
\label{LLG}
\end{align}
here $\gamma$ is the gyromagnetic ratio, $\alpha$ is the Gilbert damping, $\boldsymbol{v}_i$ is the magnon or electron velocity, 
$\boldsymbol{m}={\boldsymbol{M}}/{M_s}$ is the normalized magnetization with the magnetization vector $\boldsymbol{M}$ and  
the saturation magnetization $M_s$, $a_i$ and $b_i$ are adiabatic and diabatic coefficients of magnons or electrons, respectively
($i=m\,\,\mathrm{or}\ s$ represents the magnon or electron), $\mathcal{V}_m$ is the magnon velocity \cite{lin2014Ac}, 
and $\mathcal{V}_s$ is the electron velocity.
Among them, the effective field can be expressed as 
\begin{align}
\boldsymbol{h}_{eff} =& \frac{A}{\mu _0M_s}\nabla ^2 \boldsymbol{m}-\boldsymbol{h}_{dm}
+\frac{K}{\mu _0M_s}\hat{n} \\ \nonumber
&+\boldsymbol{h}_a+\boldsymbol{h}_e+\boldsymbol{h}_{td},
\end{align}
herein $A$ is the exchange stiffness, $D$ is the DMI constant and can be positive or negative here [i.e. the skyrmion chirality 
$D^{\pm}$ ($D^-\sim D<0 $ and $D^+\sim D>0 $)], 
$K$ is the magnetocrystalline anisotropy constant, $\hat{n}$ is the easy axis unit vector, $\boldsymbol{h}_a$ is the external 
magnetic field along the z-axis, 
$\boldsymbol{h}_{dm}=\frac{2D}{\mu _0M_s}\nabla \times \boldsymbol{m}\,\,\mathrm{or} \ \frac{2D}{\mu _0M_s}
\left[ \left( \nabla \cdot m \right) \boldsymbol{z}-\nabla m_z \right]$ is the (bulk or interface) DMI effective field,
$\mu_0$ is the vacuum permeability, $\boldsymbol{h}_e$ is the entropy equivalent field, and $\boldsymbol{h}_{td}$ is the thermal dipole field.

Then, we employ the collective coordinate transformation \cite{thiele1973SteadyStatea,thiele1974Applicationsb} 
$\boldsymbol{m}\left( \boldsymbol{r},t \right) =\boldsymbol{m}\left( \boldsymbol{r}-\boldsymbol{r}_s\left( t \right) \right)$
($\boldsymbol{r}_s\left( t \right)$ is the skyrmion centroid-coordinate, i.e. $d_t\boldsymbol{r}_s\left( t \right) =\boldsymbol{v}$)
to deal with Eq.\eqref{LLG}, and 
the total result here is (see the supplementary material for detailed derivation processes)
\begin{gather}
-\hat{z}\times (\boldsymbol{v}+\sum_{\boldsymbol{i}}{a_i\boldsymbol{\mathcal{V} }_i})+\eta (\alpha \boldsymbol{v}-\sum_{\boldsymbol{i}}
{b_i\boldsymbol{\mathcal{V}}_i})-\frac{\gamma \mathcal{F} }{4\pi Q\mu _0M_sm_ed}=0. \label{z}
\end{gather}
Rewriting Eq. \eqref{z} in the scalar form with two orthogonal directions and adding the thermal diffusion effect, we have
\begin{gather}
v_\parallel = \frac{\sum_{i}{\left(-a_i+b_i\alpha\eta^2\right)\mathcal{V}_i}+\frac{\alpha\gamma\eta}{4\pi Qd M_sm_e\mu_0}
(\mathcal{F}_\parallel+\mathcal{F}_\bot)}
{1+\alpha^2\eta^2}+\vartheta(T) G, \label{cz}    \\ 
v_\bot = \frac{\sum_{i}{\left(b_i+a_i\alpha\right)\eta \mathcal{V}_i}+\frac{\gamma}{4\pi Qd M_sm_e\mu_0}
(\mathcal{F}_\parallel+\mathcal{F}_\bot)}{1+\alpha^2\eta^2}, \label{qx}
\end{gather}
where $a_i=2SP\lambdabar_{a_i}$ and $b_i={\lambdabar_{b_i} R_s k_i}/{2\pi}$ represent the adiabatic and diabatic strengths of 
magnons or electrons by the scattering amplitude, $S$ is the spin size, $P$ is the spin polarization, 
$\lambdabar_{a_i}$ is the transmission coefficient,
$\lambdabar_{b_i}$ is the scattering coefficient, $R_s$ is the skyrmion radius, 
$k_i$ is the wavenumber of magnons or electrons, $\mathcal{F}_\parallel=\mathcal{F}_e+\mathcal{F}_{td}+\mathcal{F}_k$ is the sum of the equivalent forces, 
$\mathcal{F}_{\bot}$ is the force perpendicular to the particle-flow direction,
$\boldsymbol{\mathcal{F}}_{e}=0.9 |D| d \int(T-T_{0}) m_{e}^{4.6}(\partial_{T}m_{e})^{2}d T \hat{r}_{\parallel}$ represents the magnon entropy effect 
($d$ is the film thickness, and $T_0$ is the heat source temperature),
$\boldsymbol{\mathcal{F}}_{td} = \frac{1}{2} \pi \mu_{0} M_{s}^{2} w^{2}d^{2} G^{2} \int \frac{1}{T-T_{0}} m_{e}(\partial_{T} m_{e})^{2}dT 
\hat{r}_{\parallel}$ represents the effect of the thermally-induced dipole field
($\mu_0$ is the vacuum permeability, $w$ is the heating source radius, and G is the magnitude of thermal gradients),
$\boldsymbol{\mathcal{F}}_{k}=\frac{2 \pi K d}{G} \int m_{e}(\partial_{T} m_{e})^{2}(T-T_{0})^{2}dT \hat{r}_{\parallel}$ represents the 
thermally-induced magnetocrystalline anisotropy gradient, $\vartheta(T)$ characterizes the thermal diffusion accompanied by the thermal activation,
$\eta$ is the damping coefficient, $\mathcal{V}_m=\mu_Bj_m/em_eM_s$ is the magnon velocity \cite{lin2014Ac}, 
$\mathcal{V}_s=-\mu_Bj_s/em_eM_s$ is the electron velocity (the minus sign is because the electron-flow direction is opposite the current), 
$j_m=-\kappa_rdT/dr_\parallel$ is the magnon flow surface density, 
$j_s=\sigma S_\sigma dT/dr_\parallel$ is the current surface density \cite{wang2020Thermala}, $\mu_B$ is the Bohr magneton, 
$S_\sigma$ is the Seebeck coefficient, $\sigma$ is the electrical conductivity, and $\kappa_{r}(h_a, T)$ is the magnon thermal-Hall 
conductivity \cite{onose2010Observationa}. 

Subsequently, we clarify the different mechanisms of the individual field component. 
Firstly, the magnons once generated will immediately move to the cold region, leading to the larger magnon state density bound in skyrmions  
than the hot region, which makes skyrmions tend to move toward the hot region to balance the magnon entropy (see Fig.\ref{Fig1} panel 1). 
Hence $\boldsymbol{h}_e$ is caused by the magnon entropy \cite{wang2014Thermodynamica} originated from the exchange 
stiffnesses varying with the temperature. 
On the other hand, skyrmions will diffuse to the cold region to increase the entropy themselves (Fig.\ref{Fig1} panel 3), causing a 
finite thermal-diffusion velocity. 
Secondly, the dipole field $\boldsymbol{h}_{td}$ should only exist in a particular heating configuration, e.g., laser heating acting on the surface 
directly as shown in Fig.\ref{Fig1}. 
Otherwise the local magnetic charge cannot form \cite{moretti2017Domainc}, and the dipole field will not exist.
Thirdly, the vertical magnetocrystalline anisotropy constitutes a gradient field under temperature gradients, but $\mathcal{F}_k$ is smaller than 
other forces, so the existing systems \cite{yu2021Realspace,wang2020Thermala,litzius2020Role,mochizuki2014Thermally} have disregarded it.

\begin{figure}[t]
\includegraphics[clip=true,width=1\columnwidth]{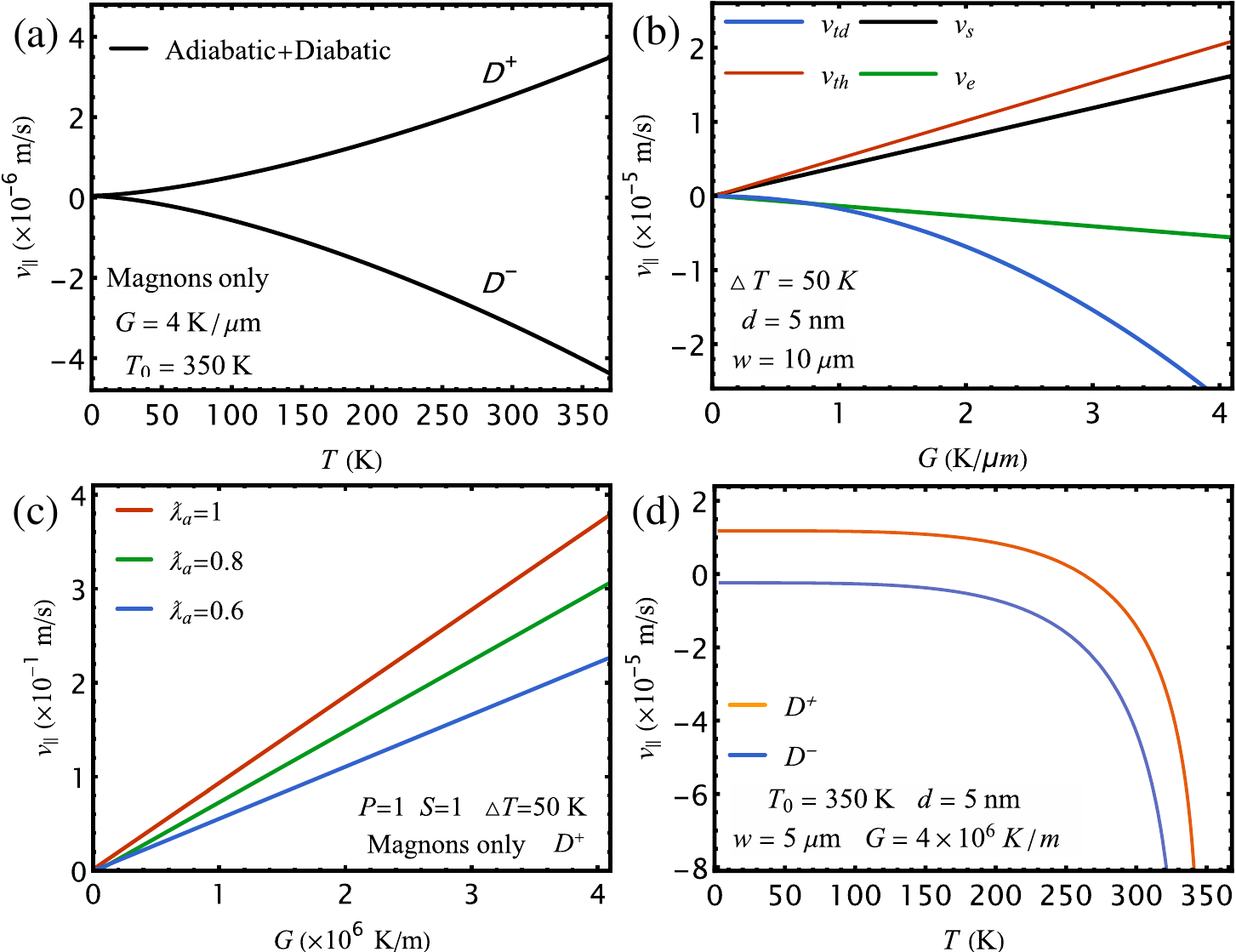}
\caption{
(a) The effect of the adiabatic and diabatic torque of magnons on $v_\parallel$. 
(b) The contribution of the adiabatic torque (s), entropy equivalent field (e), dipole field (td), and diffusion (th) to $v_\parallel$
(the electron wavenumber $k_s=3.367\times{10}^{-8}\ m^{-1}$ is very small, it has almost no diabatic effect).
(c) The dependence of $v_\parallel$ on the thermal gradient for different transmission coefficients $\lambdabar_{a_m}$. 
(d) The effect of the skyrmion chirality on $v_\parallel$.
}\label{Fig2}
\end{figure}

\begin{figure}[t]
\includegraphics[clip=true,width=1\columnwidth]{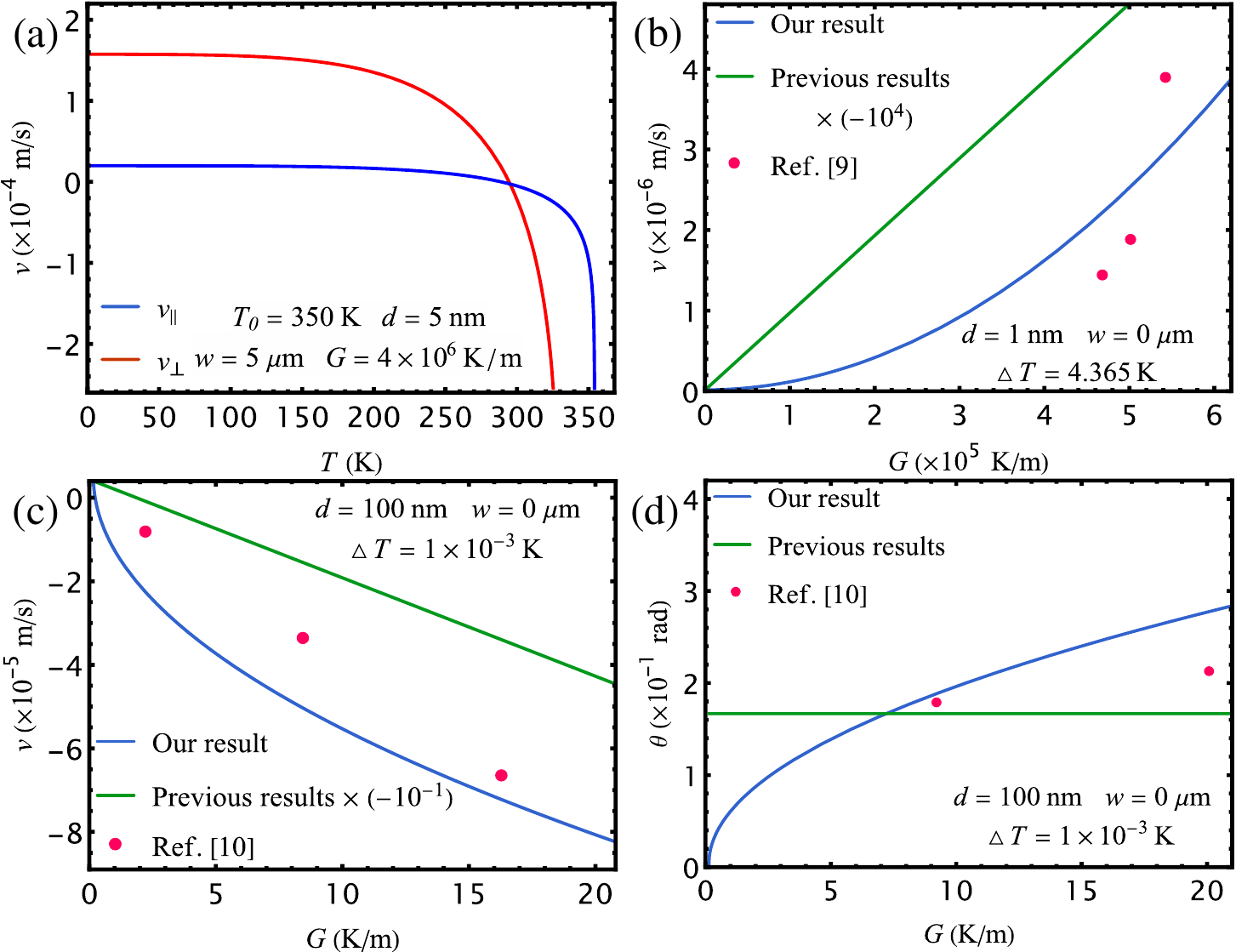}
\caption{
(a) $v_\parallel$ and $v_\bot$ vary with the temperature, and they can have the same zero crossing point.
$v_\parallel$ and $v_\bot$ can also have different zero crossing points by the variation of material parameters [not shown in the figure].
(b), (c), and (d) compare our results with experiments and existing theories.
The experimental data in (b), (c), and (d) are from Refs. [9,10] \cite{wang2020Thermala,yu2021Realspace}.
}
\label{Fig3}
\end{figure}

\begin{figure}[b]
\includegraphics[clip=true,width=1\columnwidth]{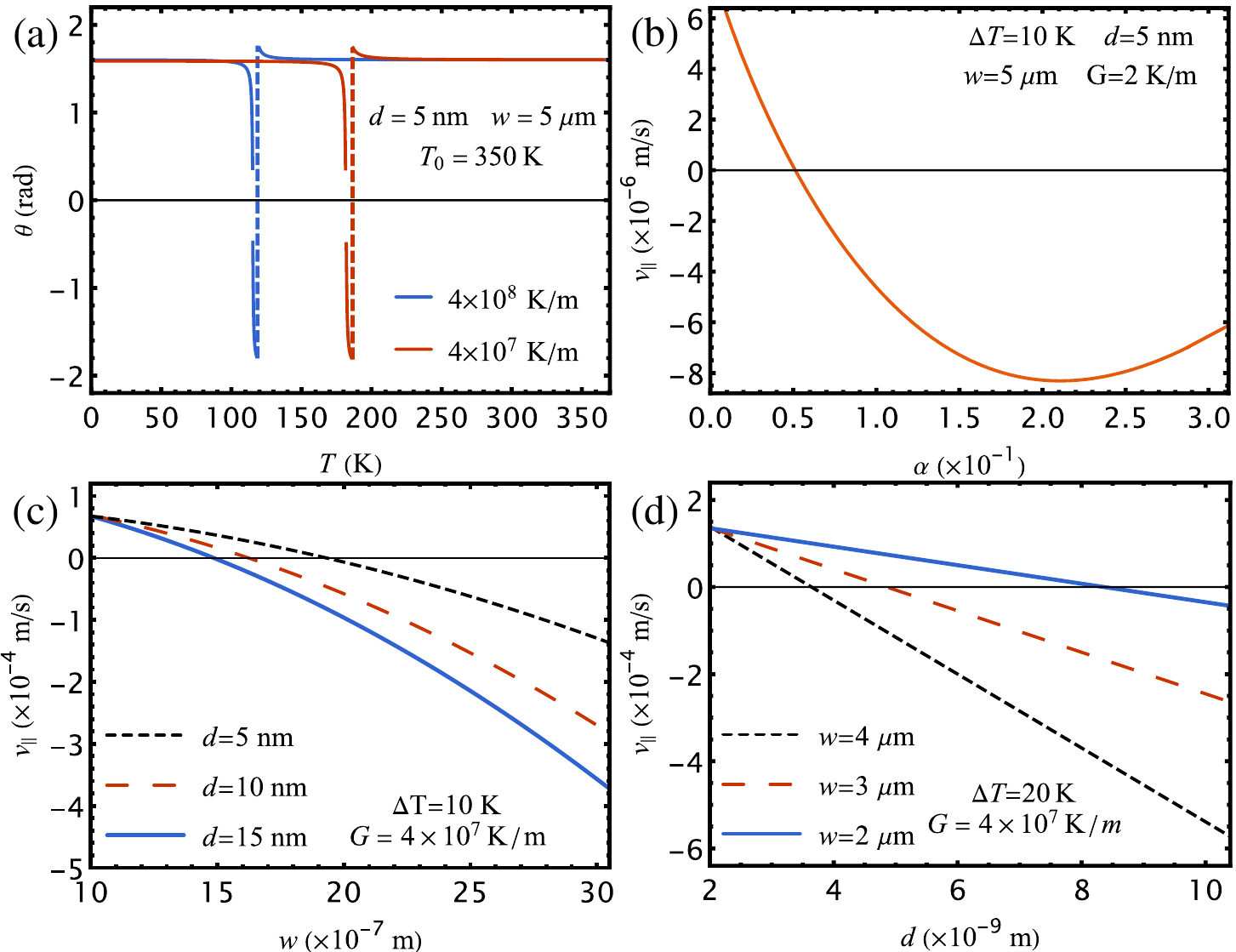}
\caption{
(a) The skyrmion Hall angle varies with the temperature under different thermal gradients. 
(b) $v_\parallel$ varies with the heat source radius $w$. 
(c) $v_\parallel$ varies with the film thickness $d$. 
(d) $v_\parallel$ varies with the Gilbert damping $\alpha$.
}\label{Fig4}
\end{figure}

In general, $v_\parallel$, $v_\bot$, and $\theta =\arctan (v_{\parallel}/v_{\bot})$ can be positive or negative due to the competition 
amongst the adiabatic moment $a$, diabatic moment $b$, entropy equivalent field $\boldsymbol{h}_{e}$, dipole field 
$\boldsymbol{h}_{td}$, and thermal diffusion [Fig.\ref{Fig2}, Fig.\ref{Fig3}, and Fig.\ref{Fig4}]. 
As shown in Figs.\ref{Fig2} (a) and (b), the effects of each component are comparable (i.e. the same order of the magnitude) under a wide 
range of experimental parameters. 
The diabatic scattering, adiabatic torque for the negative charility, and thermal diffusion can propel skyrmions from the hot to the 
cold region, whereas the entropy equivalent field, adiabatic torque for the positive charility, and the thermally induced dipole field 
will drive skyrmions from the cold to the hot region. 
The net motion of skyrmions will depend on the relative strength of these competing terms, which in turn are determined by the 
experimental parameters.   

Next, we make some comparisons with the existing theories.
Figs.\ref{Fig2} (a) and (c) indicate that the adiabatic and diabatic strength all have a significnt effect on the skyrmion velocity.  
This reveals that the widely used adiabatic and diabatic parameters \cite{zhang2004Roles} ($a_i=1$, $b_i\in(0,1)$) are only limited to the 
current-driven situation (low-frequency electrons). 
Conversely, for high-frequency incident waves (magnons), the diabatic effect can be very strong (i.e. $b_i>1$), leading to a very different 
result compared with electrons. 
Besides, the temperature also strongly affects the wave number of transmitted particles, so the universally adopted convention of 
the adiabatic limit is not well justified for finite temperatures.

Fig.\ref{Fig2} (d) shows a critical temperature which makes the skyrmion motion reverse under a proper thermal gradient,
and the trend $v_\parallel(T)$ is the same as the conclusion of Ref. [15] \cite{wang2021Rectilinear}.
Notwithstanding, they did not mark this transition because it focuses on the skyrmion motion towards the cold region.
However, their analytical expression points out the existence of this critical transition as long as appropriate parameters are selected
(this expression also supports that the total velocity of skyrmions can be zero, that is, $v_\parallel$ and $v_\bot$
have the same critical temperature [Fig.\ref{Fig3} (a)]).
They still conclude that the skyrmion Hall angle can be zero, consistent with ours, as shown in Fig.\ref{Fig4} (a).

Afterwards, we further compare our results with experiments and other existing theories. 
The experiment of M. Mochizuki et al. \cite{mochizuki2014Thermally} shows that the tangential velocity $v_\bot$ can be larger  
than the radial velocity $v_\parallel$. 
Our theory shows that $\left| v_{\bot} \right| \geqslant \left| v_\parallel \right|$ for a wide range of temperatures as shown 
in Fig.\ref{Fig3} (a), consistent with the experimental observation. 
They also show that the chirality $D$ influences on the skyrmion motion, especially when the contribution of the spin moment is 
opposite ($Q=1$, clockwise rotation, Eq.\eqref{qx}) to the relatively-obvious equivalent fields.
In our framework, the change of DMI will affect the system Hamiltonian and further impact the effective transmission and
scattering intensities.
Actually, the existence of DMI $D>0$ will enhance the reflection of particle flows \cite{yan2015Thermodynamica}, so a negative 
DMI $D<0$ should have the opposite effect.
More intuitively, this can also be explained that $D$ will induce an additional emergent field (i.e. an extra equivalent torque 
\cite{kovalev2016Spin}) and determine its direction and magnitude \cite{kim2019Tunable}.

The experiment of Z. Wang et al. \cite{wang2020Thermala} shows that skyrmions move ``abnormally'' to the cold region, in stark contrast to 
all other experiments and the existing theories [note the sign difference between Figs.\ref{Fig3} (b) and (c)]. 
The only reason for this behaviour is that the contributions of the momentum transfer ($b$) and thermal diffusion surpass the 
angular momentum transfer ($a$) and entropy equivalent field. 
Note that there is no dipole field due to the uniform heating distribution on the edge for this experimental condition. 
We use their experimental parameters to calculate the velocity as shown in Fig.\ref{Fig3} (b). 
The theory can explain the experimental observation very well whereas the previous theories of skyrmion dynamics under temperature 
gradients have not fully accomodated the interaction between transport particles (magnons and electrons) and skyrmions as well as 
the equivalent fields, incapable of explaining key experimental features (four orders of magnitudes discrepancy with the experiment).

The experiment of X. Yu et al. \cite{yu2021Realspace} shows that skyrmions move to the hot region with 
$v_\parallel \sim -10^{-6}G\,\,\mathrm{m/s}$, whereas skyrmions move towards the cold region in the experiment of Z. Wang et al.. 
In addition, a much smaller temperature gradient in Yu's experiment produces a greater velocity.
Only the variation of material parameters can not explain such a big disparity. 
In Yu's experiment, the temperature difference $\Delta T$ and global temperature are small, leading to a smaller wavenumber $k_m$ and 
skyrmion radius $R_s$, thus the adiabatic effect $a_m$ become dominant and significant (the effective force is related to the temperature 
difference).
As a result, skyrmions move to the hot region with a larger velocity [Fig.\ref{Fig3} (c)]. 
Again, our theoretical result agree very well with the experimental measurement by using their experimental parameters.
They also show that the skyrmion Hall angle $\theta =\arctan(\left. v_{\bot}/v_{\parallel} \right.)$ depends on the temperature gradient. 
It is because both components of the skyrmion velocities $v_{\parallel}$ and $v_{\bot}$ relate to the thermal gradient as indicated 
in Eqs. \eqref{cz} and \eqref{qx}.

The experiment of K. Litzius et al. \cite{mochizuki2014Thermally} (with applied currents and global-equilibrium temperatures) 
shows that the skyrmion velocity can increase with the temperature because a higher temperature heightens the skyrmion energy, 
facilitating skyrmions to cross the pinning region (thermal activation), which is shown in Fig.\ref{Fig1} (panel 4). 
Here we extend the thermal activation effect to the thermal gradient, included in $\vartheta(T)$ of Eq.\eqref{cz} 
(see the supplementary material for the detailed expression of $\vartheta(T)$).
They still show that the skyrmion Hall angle is independent of the temperature.
This can be attributed to the fact that the wavenumber of electrons is so small that there is only an adiabatic effect, and
the two velocity components of skyrmions have the same growth trend for the temperature.
So, in this case, the skyrmion Hall angle can be approximated as $\theta =\arctan(\alpha \eta)$,
which is a constant independent of the temperature and is also consistent with the previous result \cite{jiang2017Directb}.
It should be noted that, in the case of thermal gradients, the temperature in the $v_{\bot}$ direction is constant. 
Hence, the thermal creeping tends to the $v_\parallel$ direction, and Eq.\eqref{qx} has no correction factor characterizing the 
thermal activation.

Our theory can also make some inferences.
When we consider a wider temperature range and all the contributing factors, the skyrmion Hall angle will be non-monotonic as a function 
of the temperature [Fig.\ref{Fig4} (a)]. 
$\theta$ first decreases then increases and decreases again with the temperature.
The turning point of $\theta $ from positive to negative represents $v_{\bot}\rightarrow 0^+\Rightarrow v_{\bot}\rightarrow 0^-$, 
which is actually not shown in the figure because $v_{\parallel}$ is also approaching zero. 
The non-monotonicity of the skyrmion Hall angle results in the alternation of the skyrmion velocity between positive and negative values.  
Fig.\ref{Fig4} (b) shows the dependence of the parallel velocity on the damping coefficient $\alpha$.  
Unlike the previous theory \cite{kong2013Dynamics} that the velocities only increase monotonously with the Gilbert damping, it always has 
a maximum due to the non-linear effects caused by the thermal induced fields. 
Figs.\ref{Fig4} (b)-(d) show that the skyrmion velocity (both the magnitudes and direction) can be controlled by material parameters such 
as the the Gilbert damping $\alpha$, heating source size $w$, and the sample thickness $d$.

In conclusion, we have analyzed the effects of the adiabatic and diabatic momentum, entropy equivalent field, dipole field, 
thermal activation, magnetic-anisotropy gradient, and diffusion. 
We develop a unitified and hoslitic framework to describe skyrmion dynamics under thermal gradients and
address the conundrum in the literature: Skyrmions can move towards both cold and hot regions. 
Our theory can qualitatively and quantitatively explain the recent experimental results of thermal-driven skyrmion motions in 
metals and insulators. 
Some applications of thermal spintronic devices have also been proposed based on the rich and unique skyrmion dynamics driven by 
temperature gradients instead of the commonly adopted current or field driven method.

See the supplementary material for detailed derivation processes and calculation parameter values.

Chaofan Gong conceptualized and conducted the study.
Yan Zhou participated in the discussion and revised the manuscript.
Guoping Zhao discussed with C.G. and made some suggestions.

Yan Zhou acknowledges supports of the Guangdong Special Support Project (Grant No. 2019BT02X030), Shenzhen Peacock Group Plan (Grant
No. KQTD20180413181702403), Shenzhen Fundamental Research Fund (Grant No. JCYJ20210324120213037), Pearl River Recruitment Program of 
Talents (Grant No. 2017GC010293), and the National Natural Science Foundation of China (Grant Nos. 11974298 and 61961136006).
Guoping Zhao acknowledges supports of the National Natural Science Foundation of China (Grant Nos. 52111530143 and 51771127) and the 
Scientific Research Fund of Sichuan Provincial Education Department (Grant Nos. 18TD0010 and 16CZ0006).

The data that supports the findings of this study are available within the article and its supplementary material.

\nocite{*}

\bibliography{apssamp}

\end{document}